\documentclass[preprint,aps,preprintnumbers]{emulateapj}

\usepackage{natbib}
\usepackage{graphicx}
\usepackage{dcolumn}
\usepackage{bm}
\usepackage{epsfig}
\usepackage{amsmath}
\usepackage{color}
\usepackage{longtable}
\usepackage[OT2,T1]{fontenc}
\newcommand{\beq}{\begin{equation}}
\newcommand{\eeq}{\end{equation}}
\newcommand{\beqa}{\begin{eqnarray}}
\newcommand{\eeqa}{\end{eqnarray}}
\newcommand\lsim{\mathrel{\rlap{\lower4pt\hbox{\hskip1pt$\sim$}}
        \raise1pt\hbox{$<$}}}
\newcommand\gsim{\mathrel{\rlap{\lower4pt\hbox{\hskip1pt$\sim$}}
        \raise1pt\hbox{$>$}}}

\newcommand{\sini}{\sin(\rm{i})}

\begin{document}



\topmargin=16pt
\title{The {\it Posterior} Distribution of $\sini$ Values for Exoplanets
with $M_T \sini$ Determined from Radial Velocity Data}


%

\author{
Shirley Ho\altaffilmark{1,2,3},
and Edwin L. Turner\altaffilmark{4,5}
}

\altaffiltext{1}{Lawrence Berkeley National Laboratory, 1 Cyclotron Rd, MS 50R-5045, Berkeley, CA 94720}
\altaffiltext{2}{Berkeley Center for Cosmological Physics, University of California Berkeley, Berkeley, CA 94720}
\altaffiltext{3}{cwho@lbl.gov}
\altaffiltext{4}{Department of Astrophysical Sciences, Peyton Hall,
Princeton University, Princeton, NJ 08544, USA.}
\altaffiltext{5}{Institute for the Physics and Mathematics of the Universe, The University of Tokyo, Kashiwa 227-8568, Japan}

\date{\today}

\begin{abstract}
Radial velocity (RV) observations of an exoplanet system giving a value of
$M_T\sin(i)$ condition ({\it i.e.,} give information about) not only the 
planet's
true mass
$M_T$ but also the value of $\sin(i)$ for that system (where $i$ is the orbital inclination 
angle).
Thus the value of $\sin(i)$ for a system with any particular observed value 
of
$M_T\sin(i)$ {\it cannot} be assumed to be drawn randomly from a
distribution corresponding to an isotropic $i$
distribution, {\it i.e.,} the presumptive {\it prior} distribution .  Rather, the
{\it posterior} distribution from which 
it is drawn depends on 
the
intrinsic distribution of $M_T$ for the exoplanet population being 
studied.
We give a simple Bayesian derivation of this relationship and apply it to
several "toy models" for the (currently unknown) intrinsic distribution of 
$M_T$.
The results show that the effect can be an important one.  
For example,
even for simple power-law distributions of $M_T$, the median value of
$\sin(i)$ in an observed RV sample can vary between $0.860$ and $0.023$
(as compared to the $0.866$ value for an isotropic $i$ distribution) for 
indices
of the power-law in the range between $-2$ and $+1$,
respectively. Over the same range of indicies, the $95\%$ confidence interval on
$M_T$ varies from $1.002$-$4.566$ ($\alpha = -2$) to $1.13$-$94.34$ ($\alpha = +1$) times larger than $M_T\sin(i)$ due
to $\sin(i)$ uncertainty alone.  More complex, but still simple and plausible, distributions 
of
$M_T$ yield more complicated and somewhat unintuitive {\it posterior} $\sin(i)$ 
distributions.
In particular, if the $M_T$ distribution contains any characteristic mass
scale $M_c$, the {\it posterior} $\sin(i)$ distribution will depend on the ratio of 
$M_T\sin(i)$
to $M_c$, often in a non-trivial way.  Our qualitative conclusion is that 
RV studies
of exoplanets, both individual objects and statistical samples, should 
regard
the $\sin(i)$ factor as more than a "numerical constant of order unity" 
with
simple and well understood statistical properties.
We argue that reports of $M_T\sin(i)$ determinations should be accompanied
by a statement of the corresponding confidence bounds on $M_T$ at, say, the
$95\%$ level based on an explicitly stated assumed form of the 
true $M_T$ distribution in order to more accurately reflect the mass
uncertainties associated with RV studies.
\end{abstract}

\maketitle

\section{Introduction}
\label{sec:introduction}

As is well known the observational study of exoplanets began with, and
in large part has been
based on, radial velocity (RV) data which allow a measurement of the
planet's orbital parameters plus a value of $M_T \sin(i)$, where $M_T$ is
its true mass and $\sin(i)$ is the angle between the direction normal to
the planet's orbital plane and the
observer's sight line (see \cite{marcy03, marcy05, cumming08, johnson09}).  
Indeed, the very classification of an
unseen stellar companion as an exoplanet is normally made based on the 
value of $M_T \sin(i)$, hereinafter designated as $M_0$, the observed
or indicative mass.

It would, of course, be preferable to determine $M_T$ itself
and avoid the degeneracy with the largely uninteresting
random variable $\sin (i)$, and our understanding of exoplanet systems
has been greatly advanced by the relatively few cases in which the
observations of transit events allows the two parameters to be measured 
separately (see \cite{charbonneau07} and references therein and  
\cite{winn10} and references there in).

Nevertheless, the $M_T \sin(i)$ degeneracy does not seem too serious 
because it appears to be so simple and well understood.  In particular,
it seems extremely safe to assume that $i$ is randomly and
isotropically distributed or, in other words, that the orientation
of the orbital plane of an exoplanet in space is independent of the
direction from which we observe it.  However, this isotropic
distribution only describes the {\it prior} distribution of $i$,
not its {\it posterior} one, {\it i.e.,} not the relevant distribution
after it is conditioned by the measurement of an $M_T \sin(i)$ value.

%

In order to specify this isotropic $i$ {\it prior} distribution of $\sin(i)$,
consider a longitudinal strip of a sphere between $\theta = i $ and $\theta= i+d\theta$ 
in polar coordinates.  The strip 
extends around $2\pi$ in $\phi$ (the azimuthal angle), and the surface area of the strip is 
just $2\pi r^2 \sin(i) d\theta$. 
The probability of a randomly oriented vector piercing through that area is then 
just its fractional area of the whole surface of the sphere
\begin{equation}
f_i = 2\pi r^2 \sin(i) d\theta  / 2 \pi r^2 
\end{equation} 
The pdf of the inclination angle (assuming random orientation) is thus 
$\sin(i)$. In other words, the probability distribution function of the inclination 
angle of exoplanet falls into this range $i$ to $i+d\theta$ is just $\sin(i)$.


In order to determine the {\it prior} probability of $\sin(i)$, 
simply consider
\begin{equation}
f_i di = f_j dj 
\end{equation}
where $j = \sin(i)$.  After some algebraic manipulation, it is easy to show that 
$f_j$ (the pdf of $\sin(i)$ falling into a range of $\sin(i)-d\sin(i)$ and $\sin(i)+d\sin(i)$) 
is $\sin(i)/\sqrt{1-\sin^2(i)}$. 

Up to this point, the analysis is straightforward.  However, complications arise
at the next step, the derivation of the {\it posterior} distribution of $\sin(i)$,
because it depends on the {\it prior} or true distribution
of $M_T$.
As we do not 
yet know the $M_T$
distribution (see \cite{marcy05,jorissen01,udry07}), this consideration is not only an important
one in principle but might also be in practice.  The present paper is primarily intended to investigate 
this issue, the {\it posterior} distribution of $\sin(i)$ given an observation
of $M_0$, in some detail.

Before presenting a Bayesian analysis in the next sections, it may
be helpful to note that the issue resembles
familiar complications in interpreting photometric data that are
conventionally called Malmquist-type biases (see \cite{malmquist20,eddinton13,hogg98}) 
in some respects.  Namely, even
if the measurement errors are symmetric and unbiased (and, in
the simple cases most often analyzed, also gaussianly distributed...but
that is not essential), the true brightness of an astronomical object
is normally more likely to be fainter than its measured brightness than
it is to be brighter.  The well known reason is that there are
usually a larger number of fainter objects than brighter ones on
which the (symmetrical) measurement errors may act to produce the 
observed brightness.  

However, the considerations for $\sin(i)$ which we
investigate in this paper are {\it not related to measurement errors}.  It
would be unchanged even if all of the observations in question were
perfect and ideal.  Neither is it a selection bias on $\sin(i)$ of the sort 
that was briefly considered as an explanation of exoplanet RV
discoveries in their earliest days (see \cite{black97,gray97}).  Rather, we are
considering {\it the unavoidable consequences of the combination of 
a physical variable, $M_T$ with an unobservable stochastic one,
$\sin(i)$, when conditioned by a measurement of their product}.
This is, of course, a classical issue in Bayesian statistics.

Section \ref{sec:question} defines the basic question addressed by
this paper and gives a very simple illustrative example of why it
can be an important issue.  Section \ref{sec:answer} presents a
Bayesian derivation of the equations needed to answer the question
for any given distribution of masses for a population of exoplanets,
and Section \ref{sec:answer2} presents the results of the analysis
obtained by assuming various "toy models" for the true exoplanet
distribution of masses.
We then discuss observational selection effects briefly in Section \ref{sec:obs} and
conclude in Section \ref{sec:conc} with a discussion of
its practical implications for RV studies of exoplanets.
 
\section{\label{sec:question} Illustrative Example}
\label{sec:question}


The question we wish to analyze can be formulated in two equivalent but
slightly different forms, one describing the $M_T$ distribution and one the $\sin(i)$ distribution:

1) What is the probability that $M_T$ is less than $X$, given that 
RV data yield $M_0$ ($= M_T \sin(i)$)? 
The answer may be written as $P(M_T < X | M_0)$ and depends on $P(M_T)$,
the intrinsic distribution of exoplanet masses.


2) What is the probability that $\sin(i)$ is less than $Z$, given that
RV data yield $M_0$ ($= M_T \sin(i)$)?
This answer may be written as $P(\sin(i) < Z | M_0)$ and also depends on $P(M_T)$.

To relate the $\sin(i)$ probability distribution and the true mass 
distribution, it is simply: 
\begin{equation}
\label{eq:sinimt}
P(M_T < X | M_0) = 1-P(\sin(i) < Z | M_0)
\end{equation}
for $Z = \frac{X}{M_T}$.

To illustrate the fundamental issue, consider the following toy model:
Suppose that all exoplanets have a true mass of either $1.0$ $M_J$ or $2.0$ $M_J$ where $M_J$ is
the mass of Jupiter and that there are an equal number of exoplanets with each of these masses.
If an exoplanet is determined to have $M_0 = M_T\sin(i) = 0.5$ $M_J$, the value of $\sin(i)$ is 
obviously either $0.5$ or $0.25$ depending on whether it is one of the low or high true mass
exoplanets, respectively.  Moreover, since a $\sin(i)$ value of $0.5$ is about $2.236$ times more
likely than one of $0.25$ for the {\it prior} (isotropic $i$) distribution of $\sin(i)$, it follows
that the {\it posterior} distribution of $\sin(i)$ for this system consists of two $\delta$ functions,
one at $0.5$ and one at $0.25$ with the former having an amplitude $2.236$ times that of the latter.

Since any intrinsic exoplanet mass distribution could be arbitrarily well approximated by a series of $\delta$ 
functions, we can conclude that the the intrinsic mass distribution affects 
the {\it posterior} distribution of the $\sin(i)$, for any particular observed value of $M_0$. 

\section{\label{sec:answer} Bayesian Derivation of the {\it Posterior} Distributions}
\label{sec:answer}

Consider a planet at any mass $M_T$.  Given that the inclination angle
$i$ is randomly (isotropically) distributed, 
we know that (from the previous sections) the pdf of $\sin(i)$ falling into a range
of $\sin(i) - d\sin(i)$ and $\sin(i) + d\sin(i)$ is $\sin(i)/\sqrt{1-\sin^2(i)}$.
The culmulative probability of $\sin(i) < Z$ will then be

\begin{equation}
P(\sin(i)< Z)  = \int_0^Z \sin(i)/\sqrt{1-\sin^2(i)} d\sin(i)
\end{equation}

which is simply 
\begin{equation}
P(\sin(i) <Z) =  1- \cos(\arcsin(Z))
\end{equation}

This is however not surprising, since we know (from trigonometric argument) that: 
\begin{equation}
P(i<x) = 1-\cos(x)
\end{equation}

And therefore, we know that the {\it prior}  probability of finding $\sin(i)$ less than $Z$, 
which is equivalent to the {\it prior} probability of observed mass $M_0$ given $M_T$ is just

\begin{equation}
P(\sin(i) < Z) = 1- \cos(\arcsin(Z))
\end{equation}

\begin{figure}
\begin{center}
\includegraphics[width=2.5in,angle=-90]{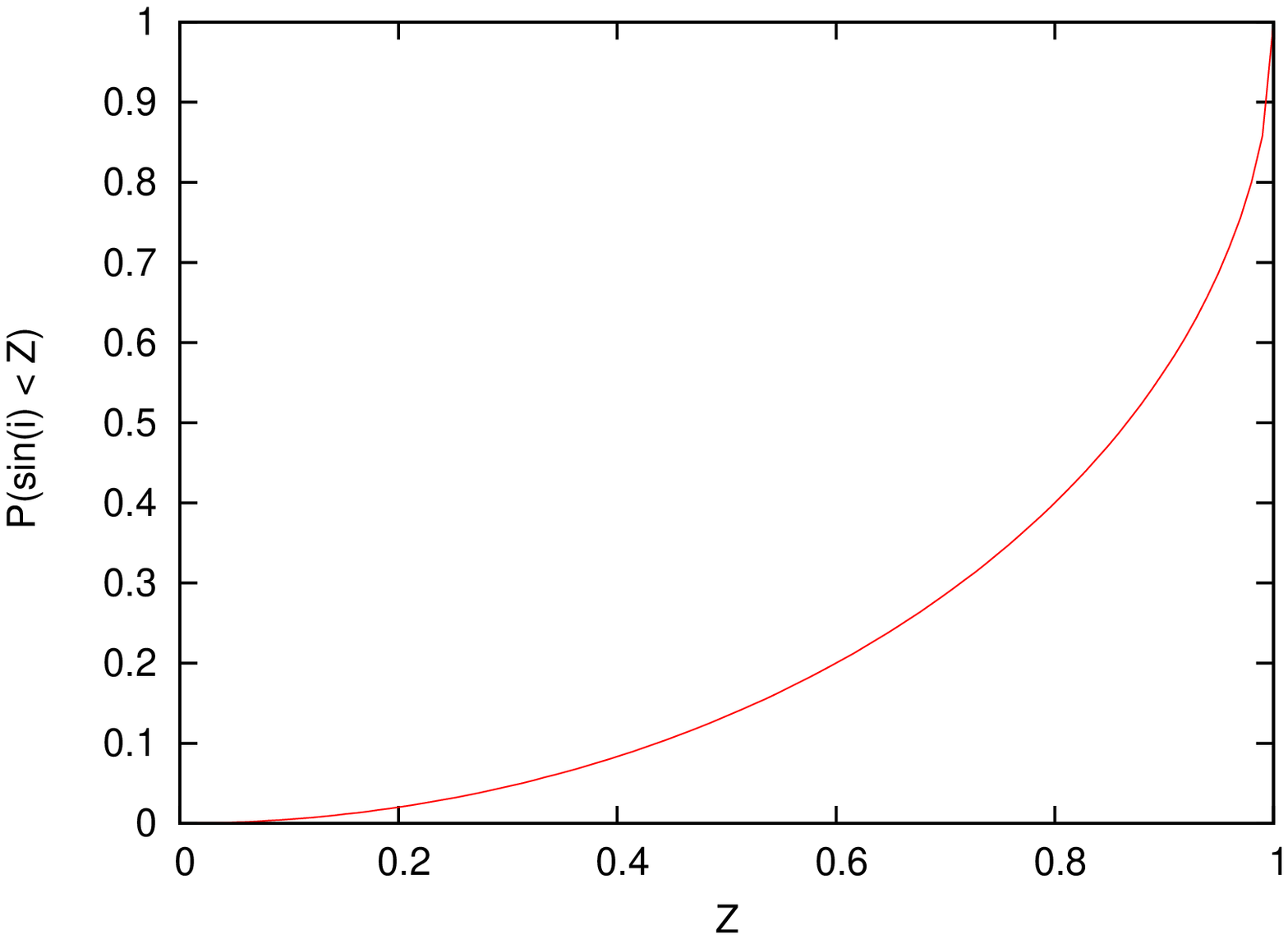}
\includegraphics[width=2.5in,angle=-90]{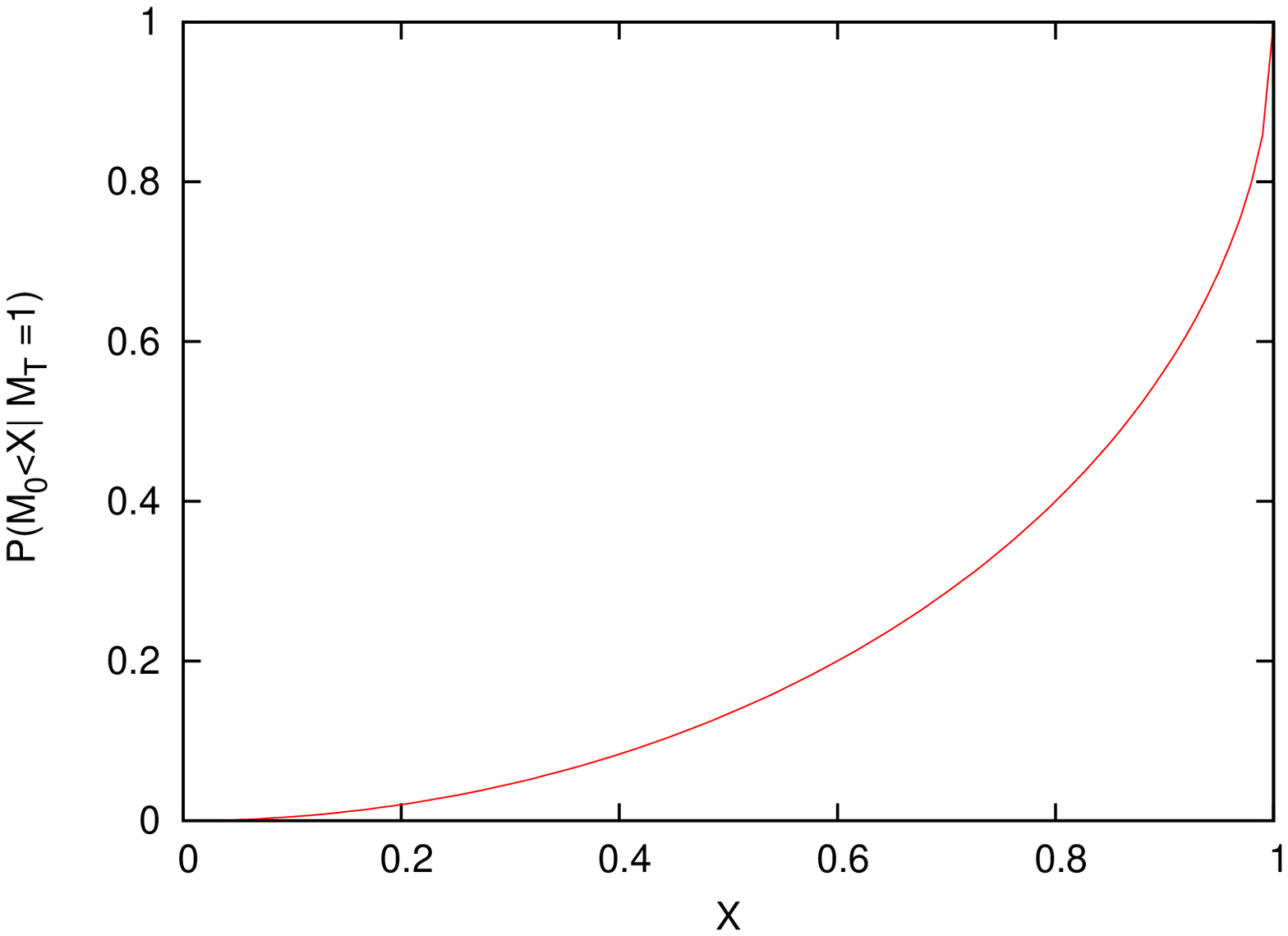}
\end{center}
\caption{The upper panel shows the cumulative {\it prior} distribution of $\sin(i)$. Note
the long tail toward small $\sin(i)$ values and thus the significant probability for a planet to have
a true mass substantially larger than its observed value of $M_T\sin(i)$. The lower panel shows the cumulative {\it prior}  distribution of $M_0$ given $M_T=1$. The two curves (as expected) are the same.}
\label{fig:m0mt}
\end{figure}

This, of course, is simply the {\it prior}
pdf of $\sin(i)$, derived in Section 1.
This distribution function is plotted in the upper panel of Fig.~\ref{fig:m0mt} in cumulative form, we can also look at the 
probability of observed mass $M_0$ given $M_T$, which is simply the following:
\begin{equation}
P(M_0 | M_T) = \frac{\frac{M_0}{M_T}}{\sqrt{1-(\frac{M_0}{M_T})^2}}
\label{eq:m0mtP}
\end{equation}

Thus, the culmulative {\it prior} probability of observed mass $M_0$ given $M_T$ (note that we are here fixing true mass $M_T$
and calculating the {\it prior} distribution of the observed mass $M_0$, while in
the analyses which follows we will do just the reverse to
obtain {\it posterior} distributions) and setting $M_T=1$ is just: 
\begin{equation}
P(M_0<X | M_T=1) = 1-\sqrt{1-X^2}
\end{equation}
As expected, the plot has the same behavior as in $P(\sin(i)<Z)$, as plotted
in the lower panel of Fig.~\ref{fig:m0mt}.

Proceeding now to {\it posterior} distributions, Bayes' Theorem states 

\begin{equation}
P(A|B) = \frac{P(B|A) P(A)}{P(B)}
\end{equation}

Assigning $A=M_T$ and $B=M_0$, we directly obtain 
\begin{equation}
P(M_T | M_0)  = \frac {P(M_0 | M_T) P(M_T) }{P(M_0)}
\label{eq:mtm0full}
\end{equation}

Equation ~\ref{eq:m0mtP}  gives the first term in the numerator of Eq.~\ref{eq:mtm0full}.
The second term in the numerator is an unknown function (which 
ultimately may be
determined from observations), but it is possible to consider simple
toy models, plausible guesses and even theoretical estimates for
$P( M_T )$ and thus explore their consequences.
Finally, the denominator of Eq.~\ref{eq:mtm0full}  can be obtained from
\begin{equation}
P(M_0) = \int P(M_0 | M_T) P(M_T) dM_T
\end{equation}

Therefore 
\begin{equation}
\label{pmtm0}
P(M_T| M_0) = \frac{P(M_0 | M_T) P(M_T)}{ \int P(M_0 | M_T) P(M_T) dM_T }
\end{equation}

which provides the desired {\it posterior} distribution 

\begin{equation}
P(M_T| M_0) = \frac{\frac{\frac{M_0}{M_T}}{\sqrt{1-(\frac{M_0}{M_T})^2}} P(M_T)}{ \int \frac{\frac{M_0}{M_T}}{\sqrt{1-(\frac{M_0}{M_T})^2}} P(M_T) dM_T }
\end{equation}

for $M_0 < M_T$.

It is frequently most interesting to consider instead
the cumulative probability distribution at which 
$M_T < X$, requiring the integration of the numerator of 
Eq.~\ref{pmtm0} up to X.  This gives

\begin{equation}
P(M_T < X | M_0) = \frac{\int^X P(M_T| M_0) P(M_T) dM_T}{ \int P(M_T| M_0) P(M_T) dM_T }
\end{equation}

Since if the observed mass is $M_0$, then the true mass $M_T$ has to be larger than or equal 
to $M_0$, since $M_0 = M_T \sin(i)$ and $\sin(i) \le 1$, therefore the lower integral limit 
is $M_0$, and the upper mass limit could be as large as physically possible for mass 
of a planet ($M_{max}$).

Therefore, we have the following: 
\begin{equation}
\label{eq:mprob}
P(M_T < X | M_0) = \frac{\int_{M_0}^X \frac{\frac{M_0}{M_T}}{\sqrt{1-(\frac{M_0}{M_T})^2}} P(M_T) dM_T}{ \int_{M_0}^{M_{max}} \frac{\frac{M_0}{M_T}}{\sqrt{1-(\frac{M_0}{M_T})^2}} P(M_T) dM_T }
\end{equation}

The upper bound on the
integral in the denominator $M_{max}$ is somewhat arbitrary,
corresponding to the maximum mass of any planet drawn from the
$P(M_T)$ distribution.  However, the value of $M_{max}$ affects only
the normalization of $P(M_T < X | M_0)$, not its form.


This formulation in terms of $M_T$ most transparently displays the
underlying logic of the derivation.  However, the same
approach can equally well
give the answer to question \#2 above, since the 
two are equivalent.  In particular,


\begin{equation}
P(\sin(i) < Z | M_0) = P(M_T > X | M_0)
\end{equation}
\newline

And, thus

\begin{equation}
\label{eq:siniprob}
P(\sin(i) < Z | M_0) = 1- \frac{\int_{M_0}^{M_0/Z} \frac{\frac{M_0}{M_T}}{\sqrt{1-(\frac{M_0}{M_T})^2}}  f_{M_T}(y)dy}{\int_{M_0}^{M_{max}}\frac{\frac{M_0}{M_T}}{\sqrt{1-(\frac{M_0}{M_T})^2}} f_{M_T}(y) dy}
\end{equation}

Due to our current ignorance of the true $P(M_T)$, we cannot evaluate 
these expressions uniquely for the actual observed values of $M_0$ of known 
exoplanets.  It is nevertheless instructive to do so for various 
assumed $P(M_T)$ distributions.  We devote the remainder of
the paper primarily to that exercise.

\section{{\label{sec:answer2} \it Posterior} Distributions for Assumed True Mass Distributions}
\label{sec:answer2}

In order to investigate the size and character of the statistical
effect under discussion, we will apply the formula derived in the
previous section to a series of simple "toy models" of $P(M_T)$,
some of which might turn out to reassemble reality at least qualitatively. 
From the previous derivation, $P(M_T)$ is assumed to be a normalized probability, however, 
in the following discussion, we will be adopting $f_{M_T}(y)$ as the absolute true mass distribution (with
no normalization). This is simply due to the fact that both of the derived Bayesian equations (Eq.~\ref{eq:mprob} and Eq.~\ref{eq:siniprob}) are normalizable quantities, so using the absolute true mass distribution $f_{M_T}(y)$ is not a problem.

\subsection{Power Law $M_T$ Distributions}

Beginning with a particularly simple possibility, we now 
assume the distribution of true
masses of 
exoplanets follow a power-law, thus we adopt the form $f_{M_T}(y) = A y^\alpha$, where
both $A$ and $\alpha$ are constants. 
Then we can evaluate the the main integral (hereafter $\Phi(M_0, \alpha) = \int \frac{\frac{M_0}{M_T}}{\sqrt{1-(\frac{M_0}{M_T})^2}} A M_T^\alpha dM_T $) of Eq.~\ref{eq:mprob} for several cases ($\alpha=1$, $\alpha=0$, $\alpha=-1$, $\alpha=-2$) to obtain
the following:

\begin{equation}
\label{eqnabc}
\Phi(M_0, \alpha) = \left\{\begin{array}{rl}
A M_T \sqrt{1-\frac{M_0^2}{M_T^2}}  & \text {if } \alpha = 1, \\
\frac{A\sqrt{M_T^2 -1}\log(2\sqrt{M_T^2-1}+ M_T)}{M_T\sqrt{1-\frac{1}{M_T^2}}} & \text {if } \alpha = 0, \\
-A \tan^{-1}(\frac{1}{\sqrt{M_T^2 -1}}) & \text {if } \alpha = -1, \\
\frac{A}{M_0} \sqrt{1-\frac{M_0^2}{M_T^2}} & \text {if } \alpha = -2.
\end{array} \right.
\end{equation}

\begin{figure}
\begin{center}
\includegraphics[width=2.5in,angle=-90]{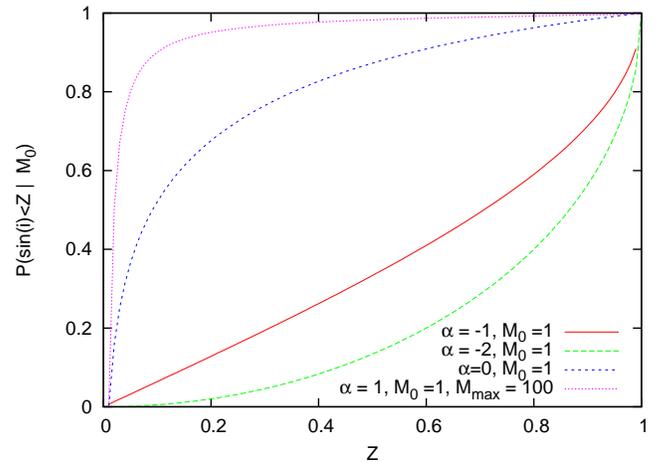}
\end{center}
\caption{The {\it posterior} probability function $P(\sin(i) < Z | M_0)$ assuming a power-law mass function. 
When we have a large $\alpha$, there are many large planets, thus, we are more likely to have 
a small $\sin(i)$ in order to match the observed $M_0$. 
When we have a small $\alpha$ or a negative $\alpha$, there are many small planets, so we 
have a higher probability of a larger $\sin(i)$ to match the observed $M_0$. }
\label{fig:alpha_sini}
\end{figure}

For the distribution of $\sin(i)$, we can refer to Eq.\ref{eq:sinimt}.
We use this result to plot a few specific cases in Fig~\ref{fig:alpha_sini}, assuming various values of $\alpha$.  

Note from Fig.~\ref{fig:alpha_sini} that $\alpha = -2$
gives a median $\sin(i)$ value of $0.860$ while an equally plausible value of $\alpha=-1$ gives a median $\sin(i)$ value of $0.704$.  
If the mass distribution is an increasing function of mass, the 
resulting median $\sin(i)$ value will be reduced quite dramatically; for example $\alpha=1$ gives a median $\sin(i)$ value of $0.02$!

\begin{figure}
\begin{center}
\includegraphics[width=2.5in,angle=-90]{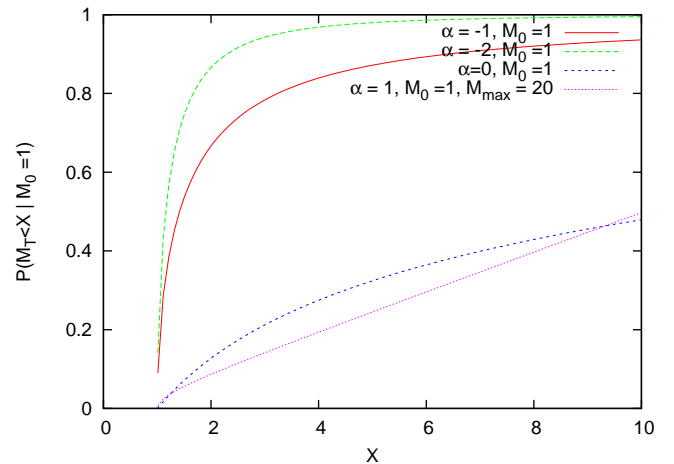}
\end{center}
\caption{The {\it posterior} probability function $P(M_T <X | M_0)$ assuming power-law mass function, with $\alpha = 1, -1, -2$, plotted with an assumption of
$M_0=1$, which is in arbitrary unit. }
\label{fig:Pm_1}
\end{figure}

It is equally easy to generate the corresponding $P(M_T < X | M_0)$
distributions, using Eq. ~\ref{eq:mprob}, as shown in Fig~\ref{fig:Pm_1}.  
Since the observed mass is set to $1$, thus, the true mass has to be larger than $1$, and 
as the power-law index increases, (which means a larger number of high mass planets in the true mass distribution), 
the probability of finding a planet below $X$ decreases (as seen in Fig~\ref{fig:Pm_1}).

\subsection{Power Law Plus a Delta Function $M_T$ Distributions}

The distribution of planetary masses in the Solar System, the highly
non-linear and at least partially non-gravitational nature of planet formation as well as some specific theoretical models (see \cite{kokubo96,kokubo06,ida08} and \cite{baraffe10} and references therein) suggest
that the $P(M_T)$ distribution might contain one or more characteristic
masses, rather than being an entirely scale free power law.  In order
to investigate the implications of such a $P(M_T)$, we consider a
toy model in which some of the
exoplanets are distributed
in an $\alpha = -2$ power law population
while the others all have the same mass $M_c$.
We may then
again evaluate the expressions of Section II directly.

Thus we have $f_{M_T}(y)= A y^\alpha + B \delta(y-M_c)$, where 
$M_c$ is the critical mass scale of interest. 
It is convenient to introduce the
dimensionless parameter $\eta$, defined by
$M_0 = \eta M_c$ and to set $A$ and $B$ equal.
Without loss of generality $M_c = 1$ is
adopted ({\it i.e.}, $M_c$ is defined as the unit of mass) for purposes
of plotting and giving numerical values.
We can then obtain the following:

\begin{equation}
P(M_T <X | M_0) = \frac{ \frac{1}{\eta}\sqrt{1-\frac{\eta^2}{X^2}} + \frac{\eta}{\sqrt{1-\eta^2}} }{\frac{1}{\eta}\sqrt{1-\frac{\eta^2}{M_{max}^2}} + \frac{\eta}{\sqrt{1-\eta^2}}}
\end{equation}
given that $M_0<  M_c < X < M_{max}$. 
It is easy to see that the addition of $\frac{\eta}{\sqrt{1-\eta^2}}$ 
will increase 
the probability that $M_T$ is smaller than $X$.

Furthermore, if $M_0< X < M_c < M_{max}$, we have
\begin{equation}
P(M_T <X | M_0) = \frac{ \frac{1}{\eta}\sqrt{1-\frac{\eta^2}{X^2}}  }{\frac{1}{\eta}\sqrt{1-\frac{\eta^2}{M_{max}^2}} + \frac{\eta}{\sqrt{1-\eta^2}}}
\end{equation}
This makes sense as the critical mass scale is not within the boundary that we consider ($M_T<X$), so the probability decreases.

Finally, if $M_0< X < M_{max}< M_c$, then the results are similar to the original situation when 
$f_{M_T}(y) = Ay^{\alpha}$ except that some of the planets are in the $\delta$
function part of the distribution, thus reducing the
relative probability of sampling the power-law portion:

\begin{equation}
P(M_T <X | M_0) = \frac{1}{2} \frac{ \frac{1}{\eta}\sqrt{1-\frac{\eta^2}{X^2}}  }{\frac{1}{\eta}\sqrt{1-\frac{\eta^2}{M_{max}^2}} }
\end{equation}

We can also obtain the distribution of $\sin(i)$: 
\begin{equation}
P(\sin(i) < Z | M_0) = 1-\frac{1}{2} \frac{ \frac{1}{\eta}\sqrt{1-Z^2}  }{\frac{1}{\eta}\sqrt{1-\frac{\eta^2}{M_{max}^2}} }
\end{equation}
if $0<\sin(i)<M_0/M_c<M_0/X$.

If we set $Z=\frac{M_0}{X}$ and $M_0 = \eta M_c$, while $M_c =1$,
then we can plot the following figure Fig~\ref{fig:sini_pow_delta}.
It illustrates the discontinuity in the probability at the delta function ({\it i.e.}, when $\eta = Z$).

\begin{figure}
\begin{center}
\includegraphics[width=2.5in,angle=-90]{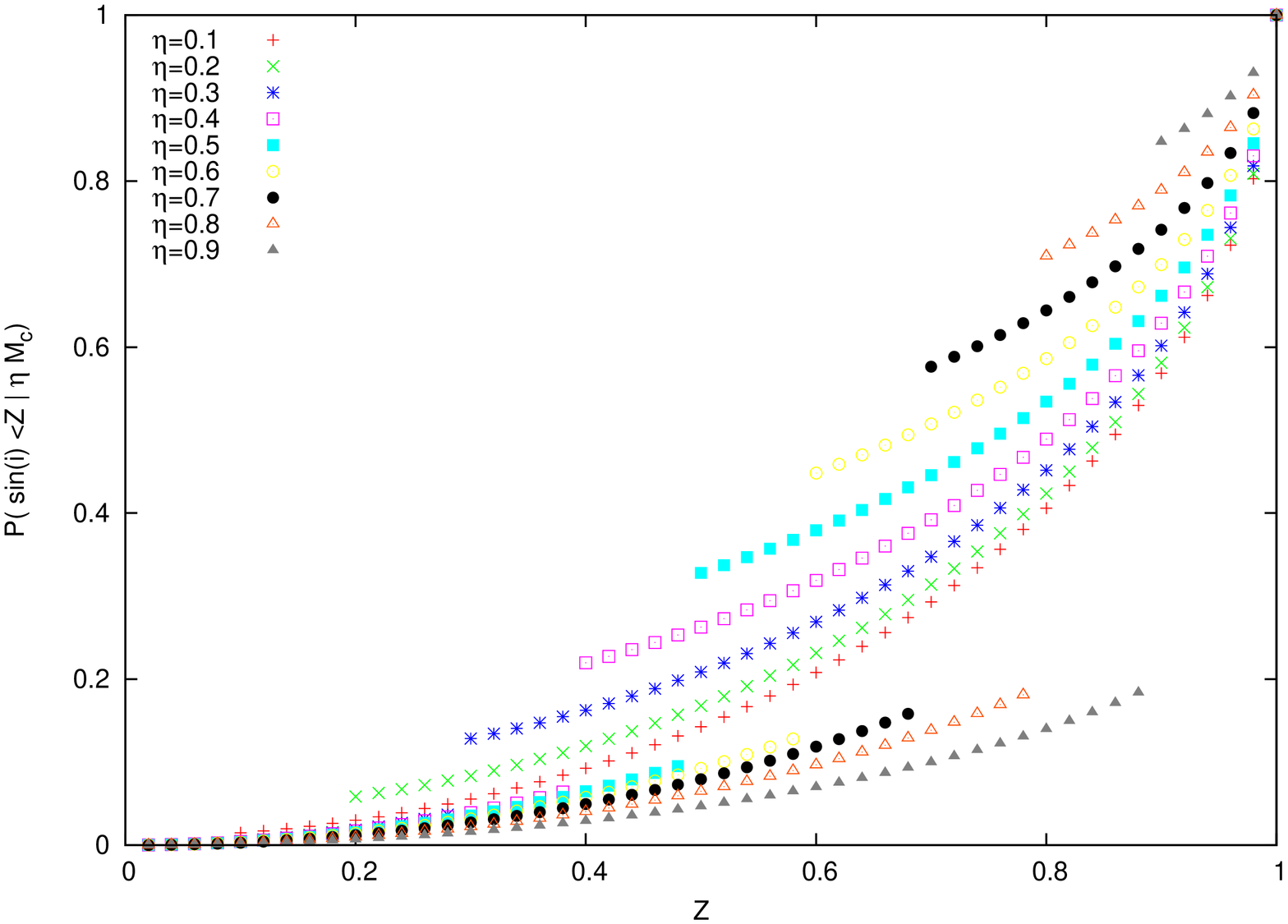}
\includegraphics[width=2.5in,angle=-90]{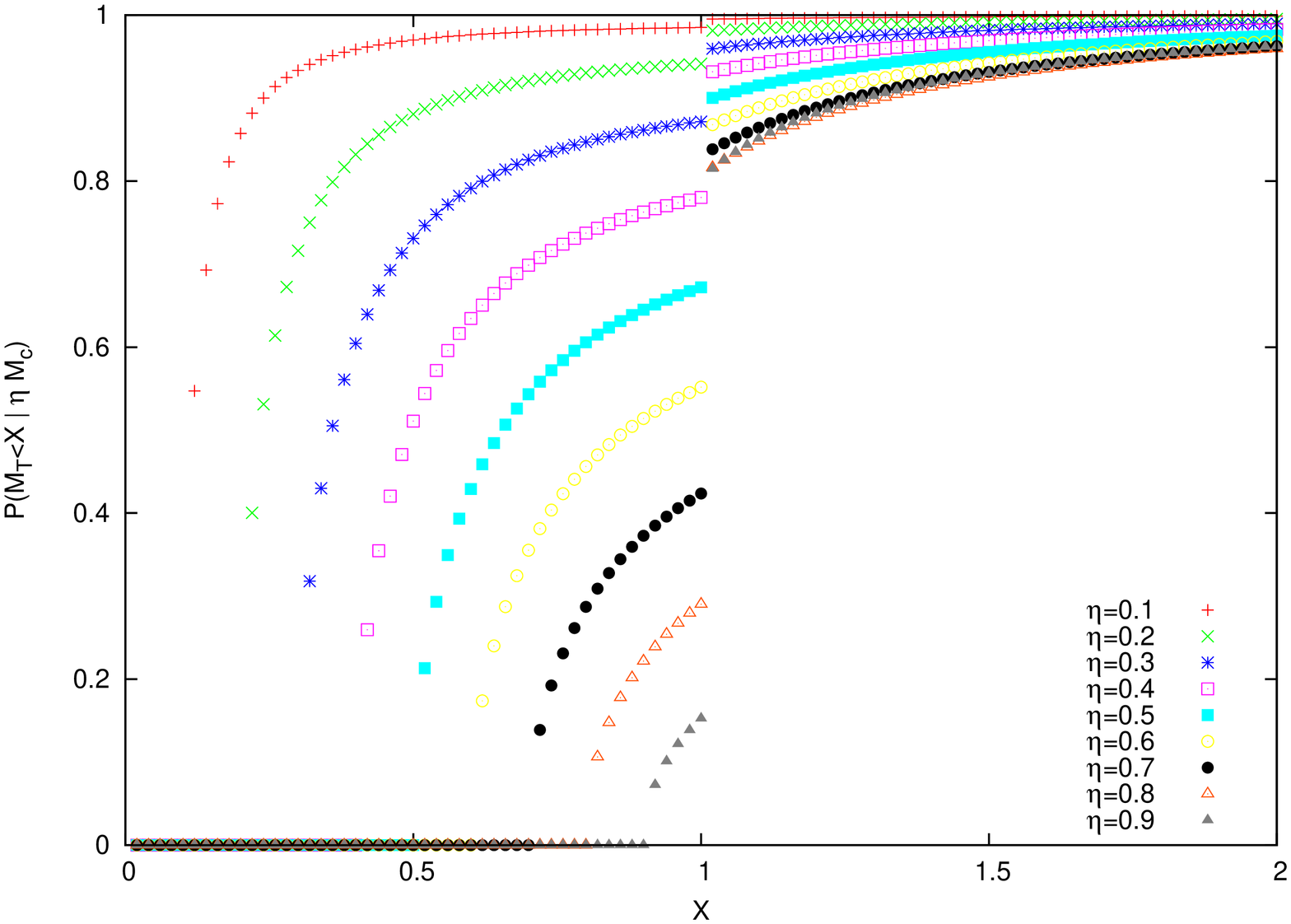}
\end{center}
\caption{The upper panel shows the {\it posterior} distribution of $P(\sin(i) < Z | \eta)$ for a power-law (with $\alpha=-2$) and 
a delta function at $M_c$. We varies $M_0$ (observed mass) with respect to $M_c$ (where the Dirac delta function is) by 
varying $\eta$ as $M_0 = \eta M_c$. One can see that there is a significant jump in probability of $P(\sin(i) < Z | \eta)$
whenever $M_0 = M_c$ (aka $\eta = Z$). The lower panel shows a similar plot, except for $P(M < X | M_0 = \eta M_c)$ as we vary $X$.}
\label{fig:sini_pow_delta}
\end{figure}

\subsection{A Solar System Like Mass Distribution}

\begin{figure}
\begin{center}
\includegraphics[width=2.5in,angle=-90]{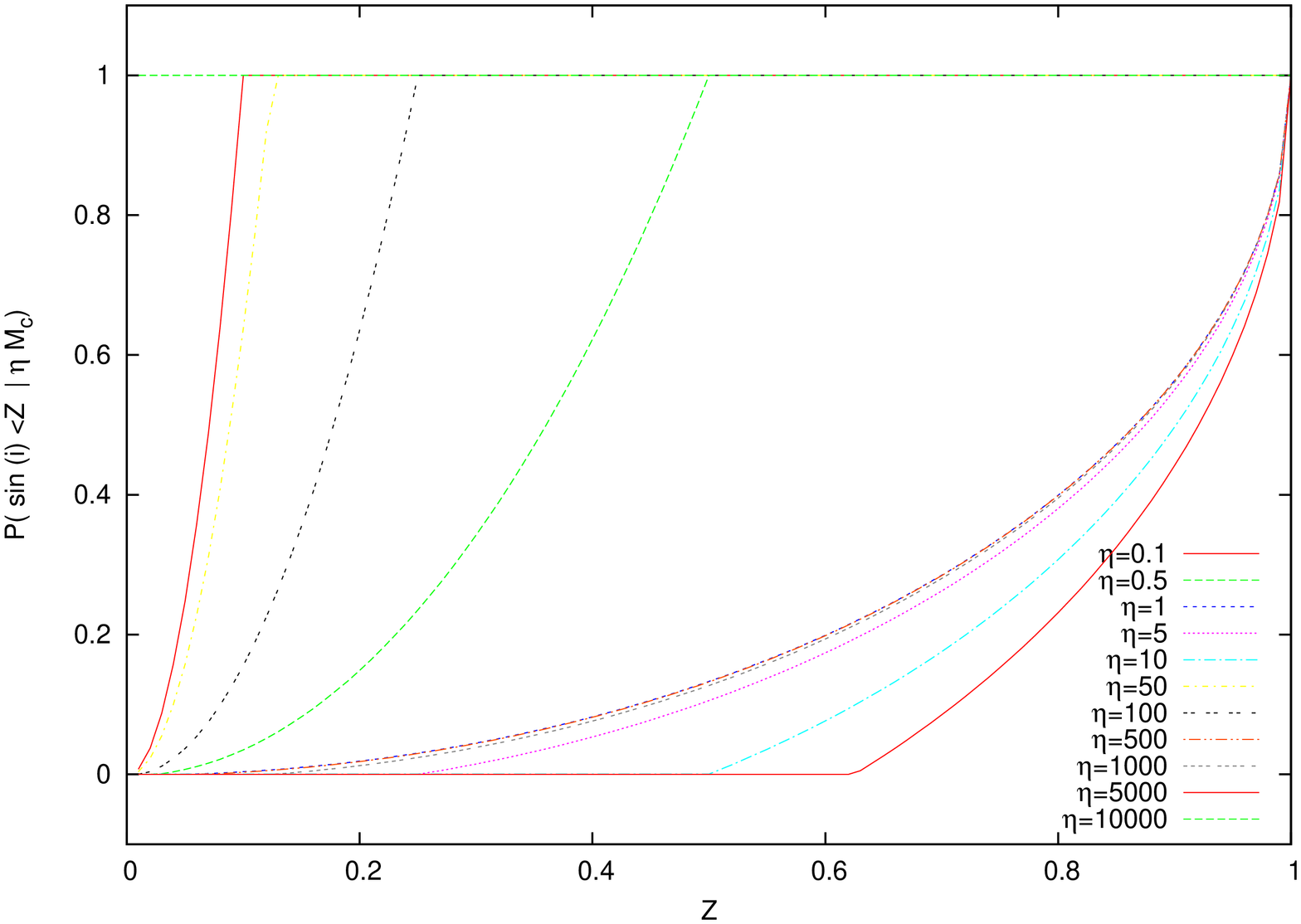}
\includegraphics[width=2.5in,angle=-90]{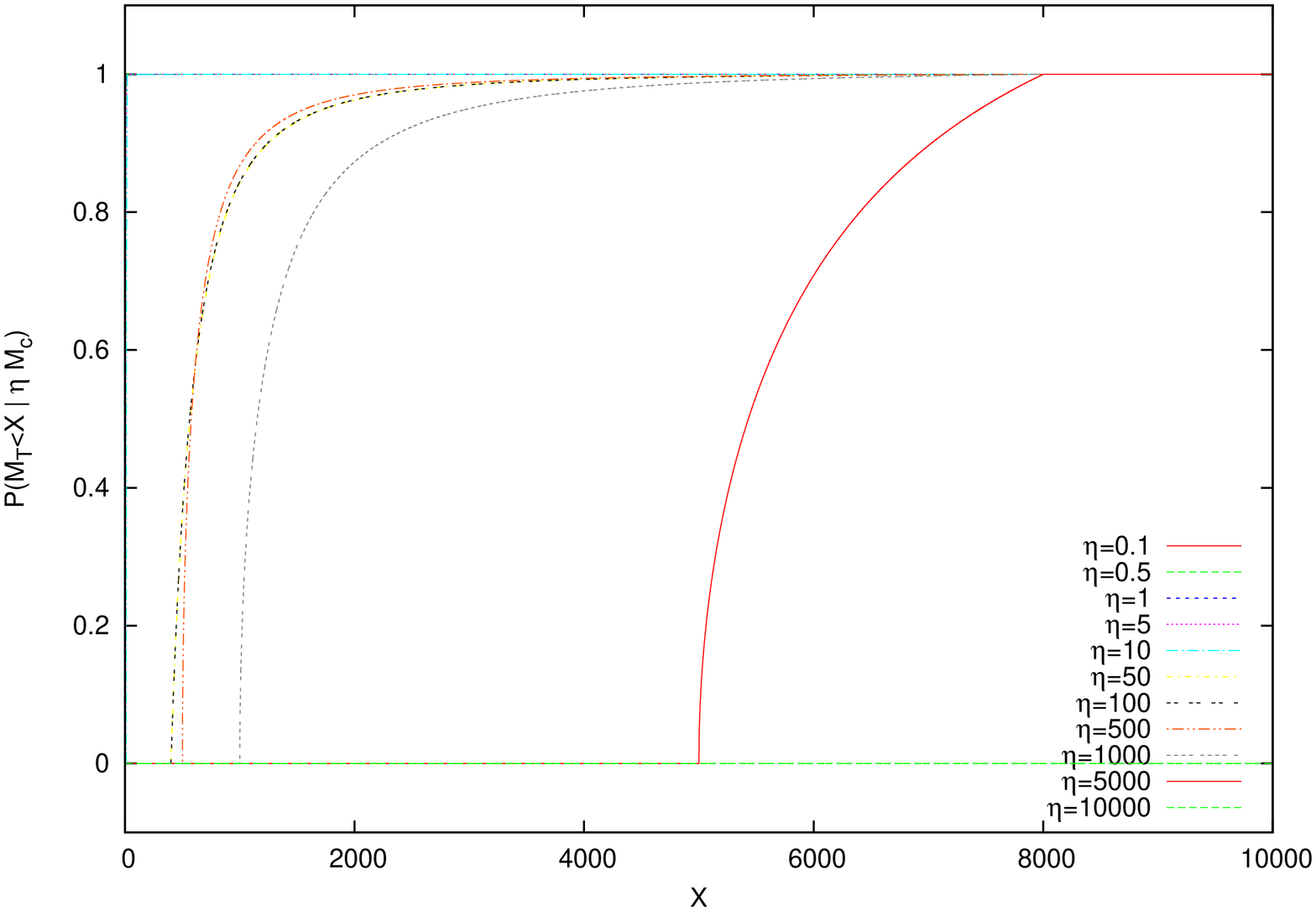}
\end{center}
\caption{Both of the above panels show the {\it posterior} probability distribution of $\sin(i)$ and $M$ for a Solar System like mass distribution. The upper panel shows the probability $P(\sin(i) < Z | \eta)$ as we change $\eta$, thus the observed mass $M_0$ goes from $0.1 M_c$ to 
$10,000 M_c$. The lower panel plots $P( M < X | M_0 = \eta M_c)$ as we change $X$. $M_c$ is set to 1 as usual.}
\label{fig:realistic}
\end{figure}

Turning now to a more complex but also
more physically plausible distribution, we analyze the case of exoplanet
masses distributed in a way similar to that of Solar System planets.
This distribution can be
modeled very roughly as two power-laws 
separated by a gap in mass. One power-law
lies at a low mass range (the
terrestrial planets) while the other lies at a much higher mass range
(the giant planets). 
We consider a toy model with 2 power-law
mass distributions, one extending from $1 M_c$ to $20 M_c$, while the other 
power-law is for $400 M_c$ to $8000 M_c$.  There are no planets in the
range between $20 M_c$ and $400 M_c$.
We also assume the two power-laws have the same power index, and also same coefficient (i.e. $f_{M_T}(y) = A y ^\alpha$ in range of $1 M_c$ to $20 M_c$ and $f_{M_T}(y) =B y^\beta$ in range of $400 M_c$ to $8000 M_c$ where $A=B$ and 
$\alpha = \beta$.)
We plot the probability $P(\sin(i) < Z | \eta)$ as $\eta$ varies (the ratio of the observed mass $M_0$ to
the critical mass $M_c$) for $\alpha = -2$ in Fig~\ref{fig:realistic}.  Note that the
probability $P(\sin(i) < Z | \eta)$ can saturate very near either unity or zero over a substantial range of $Z$ values depending on the
value of $\eta$.

\section{\label{sec:obs}Observational Selection Effects}
\label{sec:obs}

In the preceding analysis we have consistently assumed that exoplanets
discovered by the RV method uniformly (i.e., without bias) sample the
distribution of $M_T$ and $\sin(i)$ values in nature.  Obviously, this is unrealistic.  In reality, both variables (and others) influence the
probability that a given exoplanet system will be detected in an RV
survey, and this selection bias in turn affects the likely values of both
$M_T$ and $\sin(i)$.

Happily, this complication does not fundamentally alter our results
because the basic effect discussed in this paper is a purely
statistical one, independent of any observational biases.  More
specifically, one could conduct an exactly parallel analysis in which
the true distributions of $M_T$ and $\sin(i)$, which appear in equations
Eq.~\ref{eq:mprob} to Eq.~\ref{eq:siniprob}, are replaced with the biased distributions which a
particular RV survey samples, {\it if} its selection function can be
determined reasonably accurately.

A very simple example would be a case in which the probability of
an RV survey detecting an exoplanet of mass $M_T$ is given by some
selection function $S(M_T)$, independent of $\sin(i)$ and other 
properties of the system.  In that case, it suffices to replace
$P(M_T)$ with $S(M_T)P(M_T)$ everywhere it occurs in the equations
and proceed as before.




\section{\label{sec:conc} Implications}
\label{sec:conc}


The primary implication of the results presented here is that 
{\it in general the
value of $\sin(i)$ for a given exoplanet system will not be drawn from
its prior distribution, corresponding to an isotropic distribution
of i} as is often assumed, at least implicitly.

The relevant, {\it i.e., posterior}, probability distribution of $\sin(i)$ depends 
sensitively on the distribution of true masses $M_T$ and the observed
mass $M_0 = M_T\sin(i)$.
Since the former is not well constrained, either empirically or 
theoretically, at present {\it the true mass $M_T$ of such a
system cannot be trivially estimated from the value of $M_0$} as is
also often assumed to be the case (see \cite{butler04,mayor05,lovis06,wright08,mayor09}).

This means, for example, that it is difficult to identify the least 
(or most) massive RV exoplanets discovered to
date because selecting low values of $M_T\sin(i)$ from an observed
exoplanet sample is a way of picking out low $\sin(i)$ values as well
as low $M_T$ values.  For some possible exoplanet mass distributions
the observed objects with the lowest observed $M_T\sin(i)$ will be
dominated by systems with small $\sin(i)$ values rather than small
masses!

It also implies that the distribution of true exoplanet masses 
is not the same as the distribution of $M_0$ with
a constant offset in mass, corresponding to the inverse of the 
average or median value of $\sin(i)$ in the sample, as is sometimes assumed
(see \cite{mayor05,butler06,cumming08}).

The moral of the above analysis is that one should be sure to
respect $\sin(i)$ in RV exoplanet studies.
For example, we urge that RV observers reporting the value of $M_T\sin(i)$,
typically for a newly discovered planet, also report a confidence
interval for
$M_T$ at some standard selected level ({\it e.g.,} $95\%$) based on
some explicitly stated assumption for the true exoplanet mass distribution.  
 
As a simple illustration, 
the $95\%$ confidence intervals for $M_T$ if $M_0 = 1.0$ are
$1.0017$ to $4.566$, $1.005$ to $27.02$, $1.15$ to $85.186$ and $1.125$ to $94.34$
for the simple power-law $M_T$ distributions considered in Section 4.1  
with assumed
power-law slopes of $\alpha = -2, -1, 0$ and $+1$, respectively.

Although these model dependent upper bounds may appear less impressive
or exciting than the $M_T\sin(i)$ value itself, they are a less misleading
and thus more scientifically informative indication of the actual information 
on any particular exoplanet's mass provided by RV data alone.

We thank Dan Fabrycky, Scott Gaudi, John Johnson, Geoff Marcy, David Spergel,
Dave Spiegel and Jason Wright for useful comments and suggestions.
SH acknowledges support from Lawrence Berkeley National Laboratory Seaborg Fellowship 
and Chamberlain Fellowship and support from Princeton University Department of Astrophysics as SH started this project when she was a graduate student at Princeton University. 
ELT gratefully acknowledges support from a Princeton University
Global Collaborative Research Fund grant and the World Premier
International Research Center Initiative (WPI Initiative), MEXT,
Japan.



\bibliographystyle{apj}
\bibliography{msini}

\end{document}